# Computational Investigations on Polymerase Actions in Gene Transcription and Replication Combining Physical Modeling and Atomistic Simulations

Jin Yu, Beijing Computational Science Research Center

#10 West Dongbei-Wang Road, Hai-Dian District, Beijing, P. R. China, 100094

Email: jinyu@csrc.ac.cn Tel +86-10-56981807

**ABSTRACT**

  Polymerases are protein enzymes that move along nucleic acid chains and catalyze template-based polymerization reactions during gene transcription and replication. The polymerases also substantially improve transcription or replication fidelity through the non-equilibrium enzymatic cycles. We briefly review computational efforts that have been made toward understanding mechano-chemical coupling and fidelity control mechanisms of the polymerase elongation. The polymerases are regarded as molecular information motors during the elongation process. It requires a full spectrum of computational approaches from multiple time and length scales to understand the full polymerase functional cycle. We keep away from quantum mechanics based approaches to the polymerase catalysis due to abundant former surveys, while address only statistical physics modeling approach and all-atom molecular dynamics simulation approach. We organize this review around our own modeling and simulation practices on a single-subunit T7 RNA polymerase, and summarize commensurate studies on structurally similar DNA polymerases. For multi-subunit RNA polymerases that have been intensively studied in recent years, we leave detailed discussions on the simulation achievements to other computational chemical surveys, while only introduce very recently published representative studies, including our own preliminary work on structure-based modeling on yeast RNA polymerase II. In the end, we quickly go through kinetic modeling on elongation pauses and backtracking activities. We emphasize the fluctuation and control mechanisms of the polymerase actions, highlight the non-equilibrium physical nature of the system, and try to bring some perspectives toward understanding replication and transcription regulation from single molecular details to a genome-wide scale.

# I. Introduction

Polymerases are key protein enzymes that direct gene transcription and replication in the central dogma of molecular biology. They move along nucleic acid (NA) track as molecular motors [1] and catalyze RNA or DNA synthesis according to template NA strand. The chemical catalysis, mechanical performance, and fidelity control of polymerases are therefore critical for maintaining genetic health and the malfunctions leading to diverse genetic diseases [2, 3]. The polymerase enzymes are widely utilized in synthetic gene expression systems [4-6] and in genomic technologies [7-9]. Engineering and redesigning of these enzymes are highly concerned and desired for various implementations. With technological advancements in tracking and manipulating polymerase enzymes at single molecule level in recent years [10-13], fundamental mechanisms of individual polymerase actions become approachable, and that greatly improves our understandings and further implementations.

To understand the underlying functional mechanisms of polymerase enzymes with structural, dynamical, and energetic detail, computational studies are indispensible. With rapid developments on high-performance computing using parallel supercomputer clusters [14-16], molecular dynamics (MD) simulations of protein enzymes demonstrate great potential in elucidating the mechanisms from "bottom up", at a full-atom resolution [17]. Along with improvements on atomistic force field [18], long time simulations approaching microseconds to milliseconds physiological time scale have been achieved [15, 16, 19]. With improvements on sampling techniques and data analysis methods [20-24], simulations become much more efficient in evaluating energetics and other physiologically relevant observables. On the other hand, a "top down" modeling strategy toward solving specific problems at commensurable levels, as commonly practiced in physical and mathematical sciences, can effectively deal with interested properties and easily connect to experimentally measurements. In this article, we aim at providing a brief review, based on our own efforts and practices, combining both the "top down" and "bottom up" computational strategies on studying the polymerase functions.

Without being able to survey a full spectrum of computational approaches on this topic, we focus only on stochastic or kinetic modeling studies and all-atom molecular simulations that reveal mechano-chemical coupling and fidelity control properties of the polymerases. We start with a retrospect on early and general modeling frameworks built for the polymerase action, then emphasize on studies of single subunit DNA polymerases (DNAP) and RNA polymerases (RNAP) that are relatively simple in structures. For structurally more complex multi-subunit RNAPs that are studied intensively in recent years, we refer readers to two wonderful reviews [25, 26], which present highly active and detailed simulation studies on these systems. We then only show a few representative works that will shed light on future studies of polymerase transcriptional or replication controls and regulations.

## II. General physical models on polymerases

An early review on the single nucleotide addition cycle (NAC) of transcription provides a nice thermodynamics framework on RNA synthesis [27]. For each NAC, an incoming NTP is added to the existing RNA strand and the product pyrophosphate ion (PPi) is released: $RNA_i$ +NTP $\Leftrightarrow$ $RNA_{i+1}$+PPi, the RNAP also moves from position $i$ to $i$+1. The corresponding Gibbs free energy change $\Delta G$ can be decomposed into three parts: a chemical part, an RNA transcript folding part, and an RNAP elongation complex part (including the double and single-stranded DNA in the transcription bubble). In particular, the chemical part $\Delta G_{i \to i+1, chem} = \Delta G^0_{i \to i+1, chem} + k_B T \ln [PPi]/[NTP]$, with $[PPi]^{eq}/[NTP]^{eq} \sim$ 30 to 100 at an equilibrium condition [27]. The latter two parts take into account the sequence-dependent impacts from the DNA track, though on average their contributions are close to zero. This thermodynamic framework was later employed in building a sequence-dependent kinetic model of the transcription elongation [28], which made good agreements with transcription gels and single-molecule data.

Inspired by single molecule measurements on load force-velocity relationships of *E. coli* RNAP, an early mechanical model [29] treated the RNAP as a processive molecular motor capable of generating force of 25~ 30 pN. The model assumed a rate-limiting step on pyrophosphate ion (PPi) release, and suggested that NTP binding rectifies RNAP diffusion on the DNA track. Although the assumption was not confirmed by later studies, the Brownian ratchet nature of the RNAP was captured nicely in that model [29]. Based on similar single molecule measurements [30], another stochastic model of RNAP was built [31] with a focus on explaining the stall force distribution detected in the experiments. The model predicted that the stall force experimentally detected would be significantly smaller than the thermodynamic stall force [31]. In both models, DNA sequence-dependent effects had been introduced. These early modeling studies shed light on using chemical kinetics or stochastic methods to effectively describe the mechano-chemical coupling in RNAPs.

There are a few more recent modeling approaches toward understanding general RNAP properties. For example, a 'look-ahead' model for the transcription elongation has been proposed, in which NTP binds reversibly to a DNA site a few bps (~4 bp) ahead before being incorporated covalently into the nascent RNA chain [32]. The model does not concern the mechanistic nature but provides a chemical kinetic framework, in which transcription fidelity control through NTP selection is performed at several DNA template site simultaneously [32]. In another example, a general kinetic model was developed for the whole transcription cycle, taking into account that after RNA synthesis, RNAP may diffuse along DNA, desorb, or return to the promoter site to restart transcription [33]. Interestingly, the model can predict transcriptional bursts even in the absence of explicit regulation of the transcription by master proteins [33]. In a third example, dwell-time distributions in a two-state motor model was derived first, and on top of that, RNAP traffic model was developed considering steric interactions among many RNAPs moving simultaneously on the same track [34]. One more example we want to mention here is the development of a 'modular' scheme of the RNAP transcription kinetics [35], which considers alternative and off-pathway states (e.g. paused, backtracked, arrested, and terminated states) of the RNAP elongation complex. The framework can be extended to study DNA replication, repair, RNA translation etc. [35].

## III. Single subunit DNA and RNA polymerases

Below we focus on a group of single subunit polymerases [36-38], which include both DNA and RNA polymerases for gene replication and transcription, respectively. These polymerases adopt similar hand-like structures and are connected evolutionarily. We first go through studies examining mechano-chemical coupling properties of the system. These studies mainly rely on molecular modeling and simulation techniques. Then we address how fidelity control is achieved at substrate selection stage, which has been studied from both molecular simulation and non-equilibrium statistic physics perspectives.

### *III. 1 Mechano-chemical coupling in single subunit polymerases*

Since polymerases work as molecular motors, we concern about how chemical free energy is transformed into mechanical work during each NAC cycle. The chemical free energy ($\Delta G_{i \rightarrow i+1, checm}$) basically supports the phosphoryl transfer reaction, that adds the NMP part of NTP to the existing RNA strand while dissociates the PPi part. At the same time during each NAC, the polymerase undergoes substantial conformational changes to allow NTP binding and insertion; then it recovers back to the initial conformation, during or after the PPi release and translocation. Correspondingly, the mechanical motions involve both the substantial conformational changes of the polymerase and the relative translocation between the polymerase and the NA track. As a molecular motor moving along the track, the most concerned mechano-chemical coupling feature is whether the polymerase translocation is directly coupled to chemical step during the enzymatic cycle, from NTP binding to PPi dissociation. Besides, the translocation can also couple to part of the substantial conformational changes. Indeed, a previous high-resolution structural study on bacteriophage T7 RNAP suggested a power stroke mechanism [39], in which the PPi release is tightly coupled to the translocation through an O-helix or fingers domain opening motion. Under this mechanism, the PPi release energetically supports the translocation.

An all-atom MD simulation study was conducted on T7 RNAP, examining the energetics of the translocation [40]. The MD study indicated that without the fingers domain opening after the product release, the translocation is not preferred. Though large fluctuations of the RNA 3'-end were detected within the nanosecond simulations, large conformational changes and critical translocation could not be sampled. The translocation mechanism of a structurally similar DNA polymerase I (pol I) from *Bacillus stearothermophilus* was also studied by all-atom MD simulations, employing biased and targeted MD methods [41]. The study demonstrated that the PPi release precedes the translocation and facilitates the finger domain opening transition, which is then followed by DNA displacements for the translocation. Both studies suggested that the translocation of the polymerase is coupled to the opening conformational transition.

In our most recent MD studies (*ms in preparation*) on the PPi release of T7 RNAP, we constructed the Markov state model (MSM) using many nanoseconds simulations, as that performed for the multi-subunit RNAPs [42-44]. In addition, we also conducted a few microsecond simulations to detect slow motions in the release process. Interestingly, it is found that the PPi release proceeds through a 'jump-from-cavity' process, assisted by a large swing of side chain Lys472 (see **Fig 1**). The related structural features seem to be conserved in a group of structurally similar polymerases, including both RNAPs and DNAPs, so the mechanism can be general. On the other hand, the activated PPi release does not appear to be tightly coupled to the opening transition of the polymerase in the microsecond MD simulations. Hence, the studies do not support the power stroke mechanism, but are consistent with a Brownian ratchet mechano-chemical model. In the Brownian ratchet case, the PPi release precedes the translocation without direct couplings, while the translocation happens in Brownian motions without a significant free energy bias [29, 45-49]. Indeed, previous single molecule measurements on T7 RNAP only revealed a very small free energy bias toward the post-translocation state (~ 1 $k_BT$) [50, 51]. The measurements thus supported a dominant Brownian ratchet feature of T7 RNAP. Our kinetic model then suggested that the small post-translocation free energy bias could actually aid nucleotide selection in T7 RNAP [52].

From the above analyses, one can see that the substantial conformational changes in regard to the fingers domain opening and closing are crucial for the functioning of the single-subunit polymerases. Recently, the domain opening process of DNA pol I has been directly simulated using microsecond unbiased MDs at atomistic resolution [53]. An 'ajar' (semi-open) intermediate conformation, which had been discovered from a mismatched nucleotide bound structure [54], was examined in the simulation as well, and four backbone dihedrals were identified as important for the opening process.

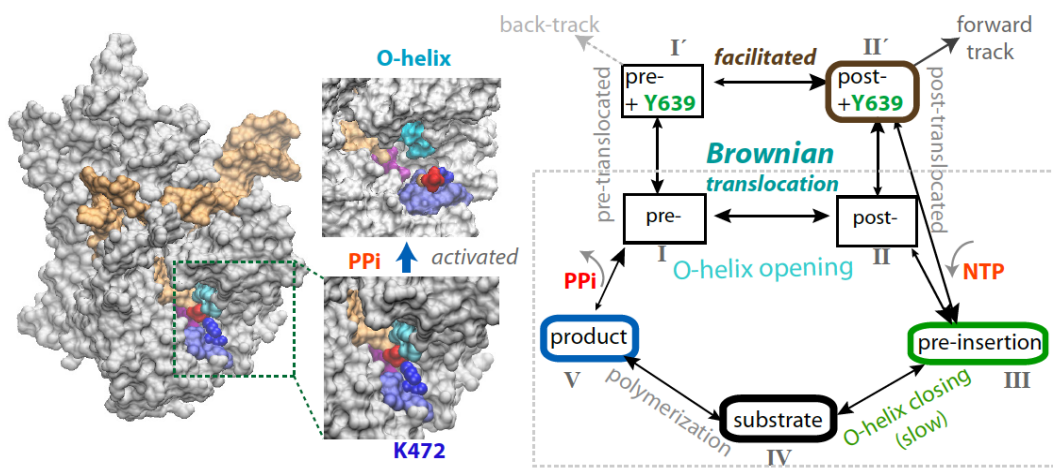

**Fig 1** Transcription elongation of T7 RNAP and mechano-chemical coupling. (Right) A kinetic scheme of T7 elongation used in a recent modeling work [52], from NTP binding (pre-insertion) and insertion (the fingers domain or O-helix closing) to the chemical reaction and product (PPi) release, followed by translocation. The translocation proceeds in Brownian movements, while the NTP binding serves for a pawl in the Brownian ratchet model to prevent backward movements. A small post-translocation free energy bias has been suggested to stabilize Y639 for incoming nucleotide selection [52]. (Left) The closed product structure of single subunit T7 RNAP elongation complex (in a surface representation: protein, while; NA, orange; the O-helix on the fingers domain, cyan; PPi, red; K472, blue, and the linked loop, light blue). The PPi release is found to be a jump-from-cavity process that is assisted by K472 side chain swing (*ms in preparation*). The release does not appear to be tightly coupled to the O-helix or the fingers domain opening, thus, cannot drive the translocation.

While the opening conformational transition after PPi release somehow couples to the translocation, the close transition after NTP binding accompanies the NTP insertion to the active site, which can be a rate limiting process in some of polymerases. The domain open/closed motion has been examined previously through elastic network models combing with normal mode analyses [55, 56]. It was noticed that the open to closed transition could be well approximated by a small number of normal modes of the open form polymerase [55]. Later, a network of residues spanning the flexible fingers domain and the stable palmdomain are found to be involved in the open-closed transition, and the conserved network of residues supports a common induced-fit mechanism in the polymerase families for the closed structure formation [56]. A comprehensive report was made recently toward understanding the pre-chemistry conformational changes in eukaryotic DNA polymerase β [57]. The NTP substrate induced domain closing transition in particular assembles the polymerase active site prior to chemistry, contributing essentially to DNA synthesis as well as on fidelity [57]. The potential of mean force for the pol β closing pathway prior to chemistry was demonstrated in the study without NTP, and in the presence of correct and incorrect NTPs [57]. It is shown that while subdomain motions appear intrinsic (as for *conformational selection*), subtle side chain motions and their favored states are largely determined by the binding of the substrate (as for *induced fit*). Hence, a hybrid of the conformational selection and induced fit mechanisms seems to apply to DNA polymerases [57].

### *III. 2 Fidelity control in single subunit polymerases*

It was generally assumed that the polymerase fidelity control is achieved through both NTP binding and chemical steps. Some of the single subunit DNAPs, such as T7 DNAP and eukaryotic DNA pol β, had been studied systematically. For example, relative stability of Watson-Crick and mismatched dNTP*template base pairs in the active site of T7 DNAP and human DNA pol β had been examined using MD simulations and linear-response analyses [58, 59]. It was found that the NTP binding selectivity of T7 DNAP is largely determined by the template-NTP interaction, while the binding contribution toward the replication fidelity control is less significant in pol β than that in T7 DNAP. Further progress understanding the fidelity control of these two types of DNAPs can be found, for example, in [60-62]. In [59], a variety of computational methods, including the free energy perturbation, the linear response approximation, and an empirical valence bond method were summarized in calculating the binding free energy contribution. In [61], the full fidelity control of T7 DNAP was studied for both the substrate binding and chemical step, by taking into account contributions from the binding, pKa shifts, PO bonding breaking and making. More recently, a binding free energy decomposition approach aiming at an accurate quantification of the pol β fidelity control was implemented [62], in which separate calculations on the neutral base and charged phosphate part using different dielectric constants were conducted.

As a small eukaryotic enzyme being able to repair short single stranded DNA, pol β has been extensively studied on its fidelity control. Beside the binding free energy, the closed to open transition of pol β was examined by targeted MD simulations in the mismatched system, in order to explain experimental results regarding inefficient DNA extension following mis-incorporation, or polymerase proofreading [63]. Recently, the crystal structures of pol β bound with the mismatched NTPs have been reported [64], together with MD simulation elucidating the replication fidelity control at both open and closed conformations. The results clearly show different DNAP responses toward different mismatches. Simulation studies on fidelity control of other single subunit polymerases also emerge recently. For example, MD studies on similar viral RNA-dependent RNA polymerases (RdRp) reveal coevolution dynamics derived from conserved and correlated dynamics of fidelity control and structural elements [65]. More recently, the RdRp from Poliovirus has been studied through MD simulations and using free energy calculation [66]. Interestingly, dynamic correlation between two important motifs appears sensitive to the incoming NTP species; the accessibility of the active site by one of the motifs also depends on the base pairing strength between the incoming NTP and the template, so that it explains why the active-site closure can be triggered by a correct NTP [66]. Furthermore, studies on HIV reverse transcriptase have ben conducted both experimentally and computationally [67]. The studies showed that the initial steps of weak substrate binding and protein conformational transition significantly enrich the yield of a reaction of a correct substrate but diminish that for an incorrect one.

Among those above studies, controversies arose, for example, on how much pre-chemistry and chemical steps contribute to the DNAP replication fidelity control [57, 68, 69]. Recently, we put up a kinetic framework on analyzing the stepwise nucleotide selection in the polymerase elongation [70], which considers contributions to the fidelity control from each kinetic checkpoint. When the elongation kinetics is described well by a three-state model (consisting of NTP binding, catalysis and translocation), the nucleotide selection can happen at two checkpoints, i.e., upon NTP binding and during chemistry step, as pointed out early. When the polymerase elongation cycle is detected with more intermediate states, however, additional kinetic transitions and checkpoints should be included. For example, NTP binding/pre-insertion can be followed by another pre-chemistry step, which then allows the NTP insertion along with an open to closed conformational transition (see **Fig 1** right). In that case, the nucleotide selection can happen at four selection checkpoints (see **Fig 2**): upon NTP binding/pre-insertion (*S1*), from NTP pre-insertion to insertion (*S2*), upon NTP insertion (*S3*), and during catalysis (*S4*). At each checkpoint, the wrong/non-cognate substrate bound the polymerase is either 'rejected' back to the previous state (*S1* and *S3*, as the wrong one faces with a lower backward barrier comparing to the right), or 'inhibited' forward toward the next state (*S2* and *S4*, as the wrong one incurs a higher forward barrier comparing to the right). The framework allows a stepwise examination of the fidelity control in a multiple-state kinetic scheme, without missing or biasing on any potential contribution.

Using the master equation approach, we demonstrated some interesting properties in the stepwise selection system. First, we notice that selection through the initial selection checkpoint (*S1*), i.e, rejecting wrong nucleotides right upon binding/pre-insertion, keeps the elongation at a relative high speed, which would not be maintained if the initial screening is not conducted.

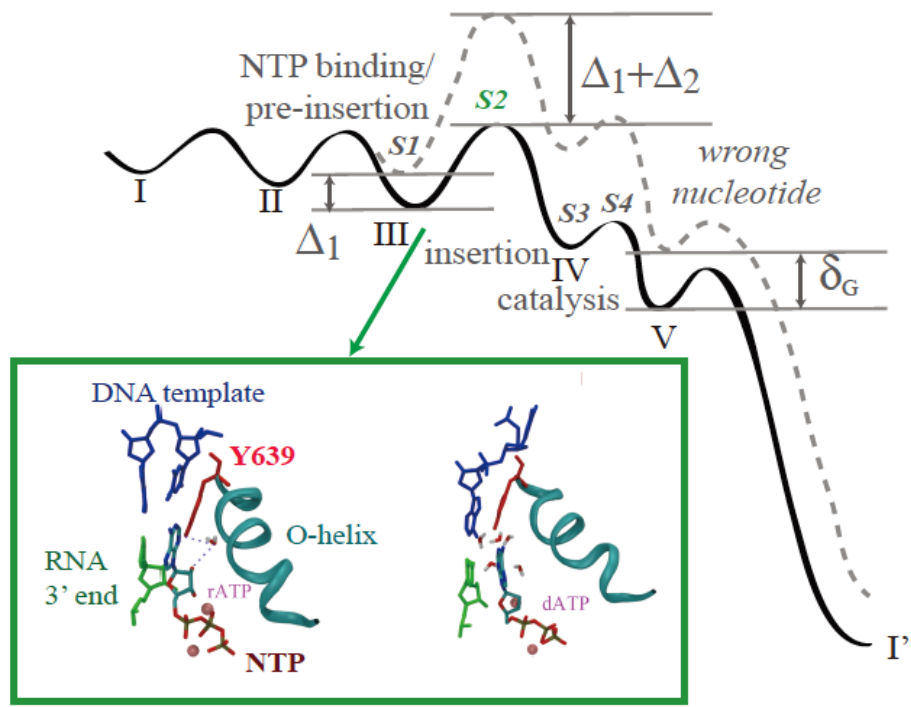

Fig 2. The nucleotide selection scheme in T7 RNAP elongation. (Top) The free energy landscape for incorporating right (solid line) and wrong (dashed) nucleotides in the five-state kinetic scheme. Four selection checkpoints (*S1* to *S4*) are labeled. $\Delta_1$ and $\Delta_2$ are differentiation free energies between right and wrong at first two checkpoints. $\delta_G$ is an overall free energy differentiation without polymerase. See [70] for detail. (Bottom) Comparing the active site configurations when the right and wrong NTP bind respectively to the pre-insertion site [71]. Y639 (red) is located on the C-term end of the O-helix (cyan) to assist the nucleotide selection. Left: rATP (right) forms the Watson-Crick base pairing with the template. The recognition is assisted by water bridging HB interactions with Y639-OH and 2'-OH of rNTP. Right: dATP cannot base pair with the template due to the Y639 interference, which associates with dATP and stacks well with DNA-RNA hybrid end, under water collision [71].

Next, we find that for a same amount of free energy differentiation $\Delta$, a same error rate is achieved for neighboring rejection and inhibition (i.e., *S1* and *S2*, or *S3* and *S4*). Finally, we show that the error rate achieved under the early checkpoints (*S1* and *S2*) is lower than that achieved later on the reaction pathway (*S3* and *S4*), if a same amount of free energy differentiation applies. One then can systematically characterize the stepwise selection i.e., by calculating the differentiation free energy at each checkpoint. We are now performing the analyses on T7 RNAP, to see if the selection system is evolved sufficiently efficient, in the absence of proofreading.

Besides, we have performed MD simulations to T7 RNAP and found a critical residue Tyr639 that assists nucleotide selection from pre-insertion to insertion [71]. This residue is marginally stabilized inside the active site in *post-translocation*, by stacking its side chain with the end bp of the DNA-RNA hybrid. A cognate rNTP (rATP, see **Fig 2** bottom) at pre-insertion site would form the Watson-Crick (WC) base pairing with the template, without further stabilizing of Tyr639 so that it can be easily pushed away during the cognate rNTP insertion. In contrast, the non-cognate would stabilize Tyr639 in the active site, so that Tyr639 keeps occupying the active site without allowing the non-cognate NTP. In particular, a dNTP (dATP, see **Fig 2** bottom) is selected against by enhancing Tyr639 stacking with the end bp, under water collision [71]. Interestingly, a non-cognate rNTP at pre-insertion grabs directly on Tyr639 instead [71]. We also studied a mutant polymerase Y639F that cannot differentiate well dNTP from rNTP, and provided molecular basis for previous experimental findings [72].

Furthermore, one should bear in mind that the polymerase elongation is a non-equilibrium process, and there are theoretical and modeling efforts made on this direction. It is understood that the equilibrium free energy difference between the right and wrong nucleotide incorporation contributes to transcription or replication fidelity, but the contribution is too small to account for the overall fidelity. The template-based non-equilibrium copolymerization process has been analyzed focusing on interplay between information acquisition and thermodynamic driving force for the copolymerization or elongation [73, 74]. It is clearly shown that the polymerase must operate far from equilibrium to achieve a high fidelity level. Interestingly, close to equilibrium, the polymerase growth or elongation can be essentially supported by configuration disorder or the incorporation of 'errors' [73]. The open-system thermodynamics to achieve DNA polymerase fidelity is systematically analyzed in [75]. In particular, the nucleotide insertion selection in the absence of the exo-nuclease proofreading had been considered. The study indicates that a sustained non-equilibrium steady state essentially drives the polymerization error rate to transit from a thermodynamically determined value to a kinetically determined one, i.e., the fidelity is achieved under the "flux-driven kinetic checkpoints" [75]. The two discrimination mechanisms involving either energetic (different binding energies) or kinetic (different kinetic barriers) differentiation are also analyzed more recently in [76]. It is shown that though the two mechanisms cannot be mixed in a single-step reaction to reduce errors, they can be combined in coping schemes with error correction through proofreading [76].

# IV. Multi-subunit RNA polymerases

Multi-subunit RNA polymerases (RNAPs) are widely distributed from bacteria to higher organisms, and have been extensively studied [77-80]. Besides those general models developed for RNAPs, early kinetic modeling and analyses were developed side by side with single molecule experiments [48, 81, 82]. For example, through a combination of theoretical and experimental approaches, a sequence-dependent thermal ratchet model of the transcription elongation was built [83]. The NTP-specific model parameters were obtained, in particular, according to the force-velocity measurements on *E. coli* RNAP [83]. A continuum Fokker-Planck framework of the RNAP elongation was also developed [84]. Using high-resolution single-molecule data [48] of *E. coli* RNAP near the equilibrium condition, a free energy profile of the polymerase translocation was obtained [84], which shows consistently the ratchet character of the RNAP elongation. Stationary distributions of the RNAP translocation at far-from-equilibrium condition (e.g. very high [NTP]) can be easily derived under this framework [84]. For multi-subunit RNAPs, we also cover two types of computational work: One is on structure-based modeling and simulations concerning molecular details of internal coupling and control. The other is on kinetic modeling focusing on backtracking, pauses, and related proofreading activities.

## *IV. 1 Probing molecular details of mechano-chemical coupling and fidelity control*

Though both high-resolution structural studies and single molecule force measurements had been extensively conducted on multi-subunit RNAPs, very detailed structural dynamics is still lack of. Nevertheless, the dynamical detail can be probed directly from 'computational microscope' at atomistic resolution. As mentioned early, systematical reviews on employing MD simulation methods to study the multi-subunit RNAPs can be found in [25, 26]. Here we only introduce representative works published very recently, which have not been included in the above reviews.

In regard to the mechano-chemical coupling of RNAP, a central concern is the translocation mechanism. The translocation is studied intensively by constructing the Markov state model (MSM) for yeast Pol II, based on a large number of short (nanoseconds) atomistic MD simulations [85]. The simulation system of Pol II reaches close to a half million atoms in explicit solvent condition. It is a big challenge, therefore, to simulate a molecular machine like Pol II up to biologically relevant time scales, i.e., from micro to milliseconds. Launching many short simulations essentially improve the computational efficiency, while constructing the MSM essentially extract the kinetic information from the simulated data. The studies show that the Pol II translocation is driven by thermal motions [85]. In particular, metastable intermediate states between the

pre- and post-translocation states have been identified. It is also found that fluctuations of a bridge helix between bent and straight conformations facilitate the translocation of the upstream RNA:DNA hybrid, which turns out to be a rate-limiting step of the translocation. The bridge helix fluctuations also facilitate the translocation of a 'transition nucleotide', which moves asynchronously from the rest of the upstream RNA and DNA in the hybrid region [85]. According to the MSM, the overall translocation rate was estimated to be about tens of microseconds at least, which is very fast comparing to the rate-limiting step of the elongation cycle (tens of milliseconds). Including the full transcription bubble may slow down the translocation rate [86]. One noted that the translocation was simulated with a trigger loop in an open conformation [85], which had been suggested to be a pre-requisite for the translocation to happen [87, 88].

Interestingly, recent single molecule experiments on Pol II identified a slow force-dependent step in the Pol II elongation, aside from a rate-limiting force-independent transition [89]. Considering that the TL opening motion can be slow and force-dependent, we built a structure-based kinetic model of Pol II elongation [90] (see **Fig 3**), attributing the force-dependent slow step to the TL opening transition prior to the translocation. On one hand, the model is made consistent with both structural dynamics studies and single molecule measurements, keeping a basic non-branched Brownian ratchet scenario; on the other hand, the model predicts the rate-limiting force-independent step conditional on accurate measurement of the NTP dissociation constant [90]: If the dissociation constant is low (high NTP affinity), then the rate-limiting step is the TL closing transition accompanying the NTP insertion; or else (low NTP affinity), the NTP incorporation transition has to be fast to avoid too much NTP dissociation, the rate-limiting step can only be the catalysis after the TL closing . The study provides a working model of the complete productive elongation cycle of Pol II, and links local structure dynamics through the non-equilibrium enzymatic cycling kinetics [90].

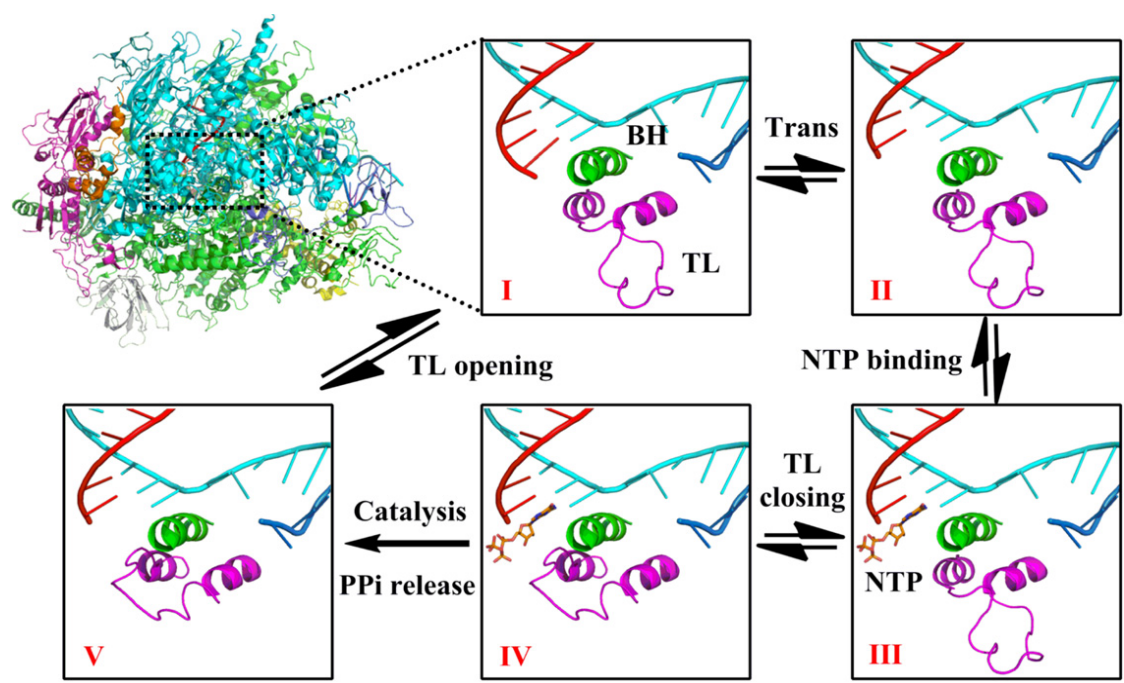

**Figure 3.** A proposed five-state Brownian ratchet model of the multi-subunit RNA polymerase II (Pol II) elongation, adopted from [90]. The structure of Pol II is provided (upper left). Configurations of the trigger loop (TL, in purple) and bridge helix (BH, in green) around the active site are shown for five kinetic states (I to V) in five windows. The non-template DNA strand is shown in blue, the template DNA strand is shown in cyan, and the synthesizing RNA strand in red. The incoming NTP molecule is shown in orange.

In regard to the fidelity control, a systematical illustration of 'five checkpoints' mechanisms is presented by using MD simulations [91]. In the multi-subunit RNAP, there is an entry site (E-site) for the NTP binding prior to the NTP insertion into the active site (A-site). Correspondingly, the five checkpoints follow the reaction path along the elongation cycle, as the initial NTP binding to the E-site, a transition or rotation of NTP from the E-site to the A-site (E-A rotation), TL closing, active site re-arrangement for catalysis, and finally, the backtracking [91]. The first four checkpoints are indeed for NTP selection, while the last checkpoint induces proofreading. In particular, the umbrella sampling method was implemented to calculate the free energy against the mismatched NTP binding at the first checkpoint, when TL is still open [91]. The studies also found that the most important checkpoint for deoxy-NTP discrimination happens when the mismatched NTP triggers conformational distortions in the active site to hinder the catalysis [91]. The final checkpoint to trigger the backtracking is through distortions of the template DNA nucleotide and DNA-RNA hybrid base pair around the active site. The studies open the door for further studies on the proposed mechanisms. It is noted, as the authors pointed out, that the efficiencies of the fidelity checkpoints on discriminating against different non-cognate rNTPs and dNTPs are sequence dependent and vary for different RNAP species.

## IV. 2 Kinetic modeling on backtracking pauses

In multi-subunit RNAPs, pauses are frequently present to play important regulation roles [92]. The paused are often linked to backtracking behaviors of the polymerases [93]. A statistical mechanics approach toward predicting backtracked pauses in bacterial transcription elongation had been conducted [94]. A thermodynamic model of the elongation complex was built with sequence-dependent free energy variations from the translocational and size fluctuations of the transcription bubble, as well as from accompanied changes in the RNA-RNA hybrid and the RNA transcript. The model produced statistically significant results toward predicting ~ 100 elongation pause sites for *E. coli* RNAP on 10 DNA templates [94]. The study also provided a kinetic model on pause recovery, assuming slow RNA unfolding and fast translocation. In both models, the sequence-specific kinetic barriers due to RNA co-transcriptional folding turn out to be essential to strongly inhibit the backtracking. Another essential feature identified was an intermediate state separating the productive elongation with the backtracking in a further developed thermal ratchet model [95]. Whether the backtracking causes a wide range of pauses, including both long and short ones, was investigated in a study later [93]. By modeling the backtracking as force-biased diffusion in a periodic one-dimension free energy landscape, the study showed a single mechanism of random walk backtracking can generate both the long (diffusive) and short (ubiquitous) pauses. In particular for short pauses, sequence-induced variations on the backward rates can have a large impact on the lifetime of the backtracking pauses [93]. Actually, when the backtracking pauses are included in a full transcription elongation model, a broad, heavy-tailed distribution of the elongation time has been obtained [96]. Interestingly, the authors of the study suggested that the pauses could even lead to bursts of mRNA production and non-Poisson statistics of mRNA levels [96], thus, contribute significantly to noise productions on a cellular level. Using a similar kinetic model and the master equation approach, these researchers studied proofreading activities involving the backtracking and RNA cleavage [97]. Backtracking by more than one nucleotide provides a multiple-checking reaction to probe the fidelity of newly generated nucleotides before further nucleotide addition. The study showed that the accuracy improves along with longer delay caused by the backtracking and cleavage. In an extreme case, the error fraction scales exponentially with the maximum backtracking distances [97]. The model thus predicts a strong dependence of transcriptional fidelity on the backtracking rates or probabilities.

## V. Summary and Perspectives

Gene transcription and replication are directed by RNA and DNA polymerases through enzymatic cycles, hence, their elongation processes are maintained at non-equilibrium steady states (NESS) driven by the chemical potential [98]. It is key to understand the NESS basis in order to understand the mechano-chemical coupling mechanisms and fidelity control features of the polymerases. The non-equilibrium statistical physics in regard to corresponding heat production, growth rate, internal entropy, and durability of the 'self-replication' process has been built up in [99]. Nevertheless, close-to-equilibrium properties of each kinetic intermediate state in the elongation cycle can be well probed through regular MD simulations, so that local structural dynamics and energetics reveal with substantial detail. For rate-limiting transitions in the elongation cycle, however, commensurable simulations should take into account the NESS chemical potential by simulating sufficiently fast processes of substrate binding and product release. The NESS dynamics would then become more of a concern when micro to milliseconds MD simulations become routine for polymerase machinery.

It is quite interesting to notice that polymerases have been largely identified to work under the loosely coupled Brownian ratchet scenario, no matter for the single or multi-subunit polymerases. In addition to the experimental evidence mentioned early, a very recent example is on translocation of replicative DNAP from bacteriophage phi29 [100]. Under the Brownian ratchet mechanism, the translocation of the polymerase spontaneously happens without being directly coupled to chemical transition such as the substrate binding or product release. However, the translocation can still couple to some essential conformational changes (such as the O-helix or TL opening in the single and multi-subunit polymerases, respectively), which may be facilitated by the chemical transition but not at the same time as during the coupling to the translocation. In contrast, the tightly coupled power stroke scenario requires simultaneous coupling between the translocation and the chemical transition, no matter other conformational changes involved or not. The power stroke scenario, however, had not been gained continuous experimental support. Besides for the polymerases, ribosomes, the most essential translation machinery, have also been consistently demonstrated to work under the Brownian ratchet scenario [101-103]. Since both machineries appeared very early in the molecular evolution history, one would speculate that the Brownian ratchet requires no highly sophisticated internal coupling mechanisms, therefore, might be easily adopted into those ancient molecular machineries.

The fidelity control of polymerase transcription and replication is achieved in general by combining nucleotide selection *before* the catalysis with proofreading cleavage *after* the catalysis. Both mechanisms work at non-equilibrium or driven conditions. The selection proceeds *stepwise* through each kinetic intermediate state, starting right after the nucleotide binding or pre-insertion, and working all the way until the end of the catalytic reaction. Substantial selection has been found to happen through the slow process of nucleotide insertion or catalysis [57, 69]. We notice that early selections outperform the late ones on the reaction path in reducing the error rate while the initial selection or screening is particularly helpful to maintain the elongation speed high. Since the template-based polymerization relies primarily on the WC base pairing, the differentiation between incoming rNTP and dNTP becomes highly subtle, involving delicate residue coordination such as 'steric gate' or hydroxyl-water interaction etc. [71, 104]. On the other hand, tolerance on template backbone sugar heterogeneity is revealed as well for Pol II [105]. There has also been evidence that the WC hydrogen bonding is not highly crucial for the fidelity control of T7 RNAP while the steric effect can be significant [106]. Remarkably, it is reported that transient WC-like mispairs (with probabilities $10^{-3}$-$10^{-5}$) stereochemically mimic the WC geometry so that to evade fidelity checkpoints [107], which can play some universal role in gene mutation and molecular evolution. One may hypothesize that the transient WC-like mispairs set a limit on the polymerase fidelity control. Anyhow, it remains elusive how much the polymerases contribute energetically to select cognates over non-cognates, especially, for variant nucleotide species. Hence, it is still hard to quantitatively test the above hypothesis.

The original idea of kinetic proofreading traced back to work of Hopfield [108] and Ninio [109]. The proofreading activities of RNA polymerases have been investigated in recent years [80]. Interestingly, in a newly published modeling work it is found that the proofreading supported fidelity control strongly depends on sequence context such that it brings to accuracy variation to several orders of magnitude [110]. Though experimentally measured free energies of dsDNA and RNA-DNA hybrid has been incorporated into the model, the polymerase contribution to the sequence-dependent accuracy variation has not been considered [110]. Again, it is because that the polymerase contribution to the accuracy has not been systematically investigated. Hence, it becomes highly desirable if computational studies in the near future could provide quantify how much the polymerases energetically differentiate cognates vs. non-cognates in a sequence specific manner. The sequence specific characterization of polymerase actions is expected to develop side by side with technology advancements on targeted and genome wide sequencing [7, 111].

# Acknowledgements

Current work is supported by NSFC under the grant No. 11275022.